\documentclass[conference]{IEEEtran}
\IEEEoverridecommandlockouts
\usepackage{cite}
\usepackage{amsmath,amssymb,amsfonts}
\usepackage{algorithmic}
\usepackage{graphicx}
\usepackage{textcomp}
\usepackage{xcolor}
\usepackage{subfigure}
\usepackage{multirow}
\usepackage{booktabs}

\def\BibTeX{{\rm B\kern-.05em{\sc i\kern-.025em b}\kern-.08em
    T\kern-.1667em\lower.7ex\hbox{E}\kern-.125emX}}
\begin{document}

\title{Content-Adaptive Rate-Quality Curve Prediction Model in Media Processing System}


\author{
    \IEEEauthorblockN{Shibo Yin$^{1*}$\thanks{$*$ These authors contributed equally to this work.}, Zhiyu Zhang$^{2*}$, Peirong Ning$^{1*}$, Qiubo Chen$^{1}$, Jing Chen$^{1}$, Quan Zhou$^{1}$, Li Song$^{2}$}
    \IEEEauthorblockA{$^1$ Xiaohongshu Inc, Shanghai, China}
    \IEEEauthorblockA{$^2$ Shanghai Jiao Tong University, Shanghai, China}
    \IEEEauthorblockA{\{yinshibo, ningpeirong, jianhan, zhouquan2\}@xiaohongshu.com, \{zhiyu-zhang, song\_li\}@sjtu.edu.cn, jingcmu@qq.com}
}


\maketitle

\begin{abstract}
In streaming media services, video transcoding is a common practice to alleviate bandwidth demands. Unfortunately, traditional methods employing a uniform rate factor (RF) across all videos often result in significant inefficiencies. Content-adaptive encoding (CAE) techniques address this by dynamically adjusting encoding parameters based on video content characteristics. However, existing CAE methods are often tightly coupled with specific encoding strategies, leading to inflexibility. 
In this paper, we propose a model that predicts both RF-quality and RF-bitrate curves, which can be utilized to derive a comprehensive bitrate-quality curve. This approach facilitates flexible adjustments to the encoding strategy without necessitating model retraining. The model leverages codec features, content features, and anchor features to predict the bitrate-quality curve accurately. Additionally, we introduce an anchor suspension method to enhance prediction accuracy. 
Experiments confirm that the actual quality metric (VMAF) of the compressed video stays within ±1 of the target, achieving an accuracy of 99.14\%.
By incorporating our quality improvement strategy with the rate-quality curve prediction model, we conducted online A/B tests, obtaining both +0.107\%  improvements in video views and video completions and +0.064\% app duration time.
\end{abstract}


\section{Introduction}
In streaming media services, it is common practice to transcode videos before distribution, effectively reducing video bitrate and lessening bandwidth pressure. A simplistic approach involves applying a uniform rate factor (RF) across all videos during transcoding. However, this often results in bandwidth wastage, as videos with minimal textural detail can achieve satisfactory subjective quality at lower bitrates, whereas videos with complex textures generally require higher bitrates to maintain acceptable quality. Consequently, researchers have proposed content-adaptive encoding (CAE), which selects different encoding parameters based on the content characteristics of each video. 
In CAE, higher bitrates are strategically increased for videos with complex content to maintain higher quality, while bitrates are judiciously reduced for simpler content to conserve bandwidth, thus enhancing overall encoding efficiency.

Netflix~\cite{aaronPerTitleEncodeOptimization2017} generates a bitrate-quality ladder by transcoding the source video at various resolutions and different RF configurations, then selects the optimal encoding resolution and RF corresponding to a particular bitrate from the convex hull of this ladder. However, this approach necessitates multiple transcoding to obtain the bitrate-quality ladder, resulting in considerable resource consumption, and rendering it unsuitable for short-form video services. With the advancement of deep learning, numerous studies~\cite{cai2022quality, covell2016optimizing, xing2019predicting, mico2023per, de2016complexity, cheng2019neural, sun2018machine, menon2023just,9858926} have employed deep neural networks to predict the optimal RF. 
For instance, ~\cite{covell2016optimizing} utilizes neural networks to predict the RF, enabling the encoder to perform constant-bitrate compression on videos with varying content. 
Alternatively, ~\cite{cai2022quality} employs neural networks to predict the RF, allowing the encoder to perform constant-quality video compression. 
However, these methods tightly couple the neural networks that predict the rate factor (RF) with the encoding strategy.
Consequently, any change in the encoding strategy necessitates retraining the model, which is quite inconvenient.

In this paper, we propose a rate-quality curve prediction model decoupled from the encoding strategy, capable of simultaneously forecasting the RF-quality and RF-bitrate curves, and subsequently deriving the bitrate-quality curve. This enables flexible adjustments to the encoding strategy based on this bitrate-quality curve. To accurately predict the rate-quality curve, the model must perceive information such as the texture, motion, and complexity of different video contents. Therefore, we propose utilizing codec features, content features, and anchor features as abstract representations of video content and predicting the rate-quality curve from these features. 
Additionally, we introduce an anchor suspension method based on anchor features to enhance prediction accuracy. 

Our main contributions can be summarized as follows:
\begin{itemize}

\item We propose a precise rate-quality curve prediction model decoupled from the encoding strategy, enabling flexible adjustment of the encoding strategy.

\item We design a video feature set containing codec features, content features, and anchor features to predict the rate-quality curve.

\item We introduce a novel anchor suspension method based on anchor features to further improve prediction accuracy.

\end{itemize}

\section{Methodology}
\label{sec:method}

\begin{figure}[htb]
    \centering
    \includegraphics[width=0.48\textwidth]{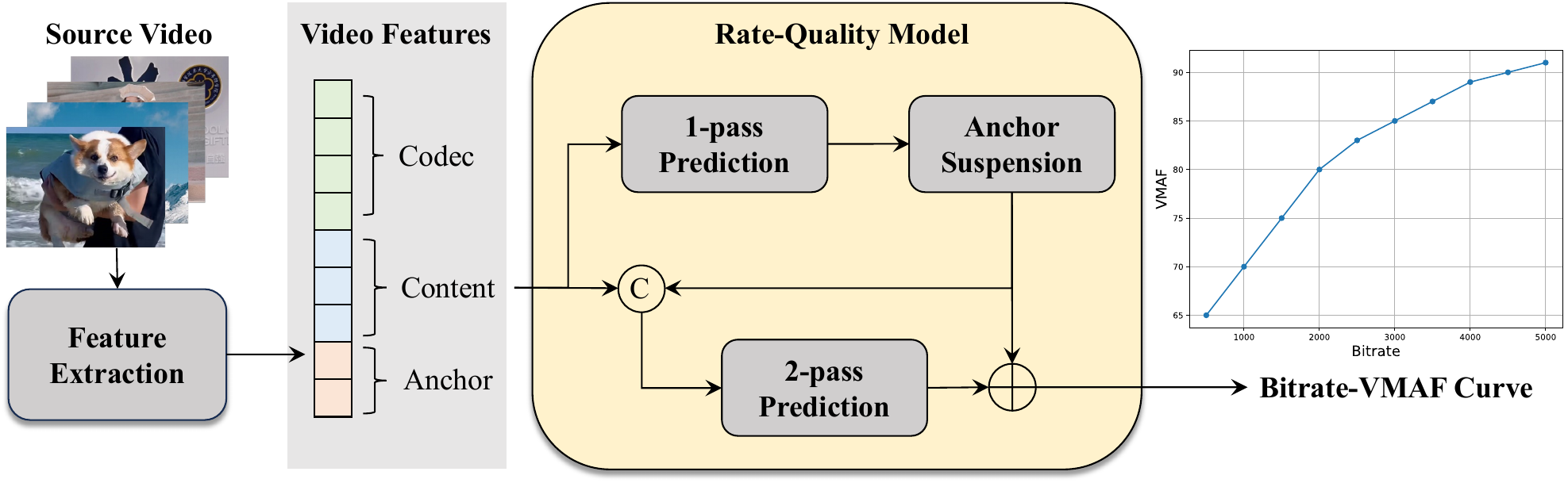}
    \caption{The pipeline of the rate-quality prediction model.}
    \label{fig:pipeline}
\end{figure}

In this paper, we construct a rate-quality curve prediction model for red265 encoder, which is based on the HEVC~\cite{sullivan2012overview} standard. The red265 encoder commonly employs the CRF parameter as RF to adjust the balance between encoding bitrate and quality. Regarding the quality metric, we utilize VMAF~\cite{li2018vmaf}, which aligns well with human perceptual assessment. We construct the rate-quality model for red265 by predicting the CRF-VMAF curve and the CRF-bitrate curve. Predicting continuous curves is a rather challenging task, we thus discretize the continuous CRF values, taking 101 discrete CRF values within the range of $[20, 40]$ with a step of $0.2$, and use the corresponding bitrate and VMAF as training data. The model solely predicts the bitrate and VMAF for these 101 CRF values.

\subsection{Overall Pipeline}

The overall pipeline is illustrated in Fig.~\ref{fig:pipeline}. First, the feature extraction module extracts video features from the source video, and then the rate-quality model predicts the rate-quality curve.
The rate-quality model comprises three stages: 1-pass prediction, anchor suspension, and 2-pass prediction. During the 1-pass prediction, a neural network predicts 101 VMAF values and 101 bitrate values based on the video features. After obtaining the 1-pass prediction results, we perform anchor suspension on the predicted curves using a prior anchor point. In the 2-pass prediction, we concatenate the suspended results with the video features as input and utilize another neural network to predict the residuals, then add the residuals to the suspended results to obtain the final output.

\subsection{Video Features Extraction}

Despite the remarkable feature extraction capabilities of CNNs in visual tasks~\cite{jiang2023mlic, zhang2024efficient, li2023neural, jiang2023slic, zhang2023high, jiang2024llic, li2024neural}, the complexity of using CNNs to extract features from videos is exceedingly high. Therefore, a lightweight approach to extract video features is required. We design three different types of features to represent the characteristics of the video: codec features, content features, and anchor features.


\vspace{2mm}

\noindent{\textbf{Codec Features.}}
Considering that the encoding information of the encoder can reflect the texture, motion, and complexity of the video, we utilize the ligthweight encoder to pre-encoding the source video and extract the encoding information. The output of each encoder can reflect the texture of the video, hence we select the faster x264~\cite{wiegand2003overview} encoder to extract the features. To further reduce computational complexity, we downsample the source video to 360p for pre-encoding. 
We select codec features based on the x264 encoding process, including the proportion of I/P/B frames, PSNR/bitrate/VMAF, block partitioning information, and prediction mode proportions, amounting to a total of 113 dimensions.

\vspace{2mm}
\noindent{\textbf{Content Features.}}
Content features reflect the inherent characteristics of the video. We derive the spatial textural features of the video based on the Grey-Level Co-occurrence Matrix, and extract the temporal correlation features utilizing conventional temporal domain algorithms. Additionally, we obtain video quality features, including noise information, no-reference quality assessment data, blockiness, and blur, from the quality evaluation module. Furthermore, we augment these features using statistical properties such as mean and variance, resulting in a 65-dimensional content features.

\vspace{2mm}
\noindent{\textbf{Anchor Features.}}
To improve prediction accuracy and algorithm stability, we introduce anchor features. Specifically, we utilize red265 to encode at a prior anchor CRF (i.e. 30.4), and obtain the actual bitrate and VMAF as the anchor features (2-dimensional). The anchor features not only serve as inputs to the neural network to assist in prediction, but also anchor the predicted curve during anchor suspension. Although acquiring the anchor features requires once additional encoding, increasing the computational complexity, the experimental results demonstrate that incorporating the anchor features significantly improves the prediction accuracy. The introduction of anchor features ensures the robustness of the rate-quality model even when the encoder undergoes minor modifications, which will be discussed in detail in Sec.~\ref{sec:anchor}.

\begin{figure}[htb]
    \centering
    \includegraphics[width=\linewidth ]{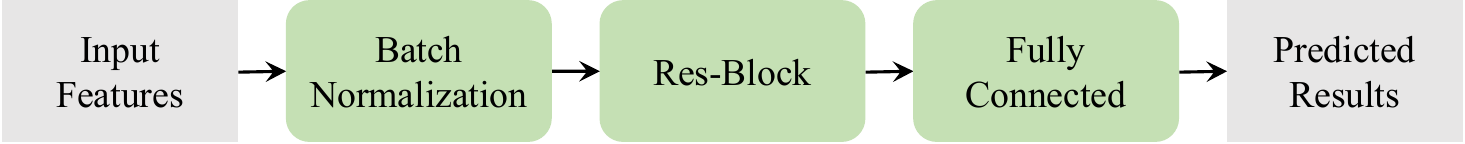}
    \caption{Illustration of neural network architecture.}
    \label{fig:architecture}
\end{figure}

\subsection{Neural Network Architecture}
We utilized the same neural network architecture for both the 1-pass prediction and the 2-pass prediction. The neural network architecture is composed of batch normalization, attention module, and residual blocks. Each dimension of the input features exhibits distinct distributions; for instance, the VMAF in the codec features ranges between $[0, 100]$, whereas some content features fall within the $[0, 1]$ interval. The disparity in feature distributions hinders the convergence of the neural network. To overcome this obstacle, we employs batch normalization to eliminate the distribution discrepancies between features and enhance training performance. Subsequently, we integrate the commonly used backbone attention module and residual blocks to enhance the model's prediction accuracy. Finally, we utilize fully connected layers to aggregate all features and output the predicted results.

\subsection{Anchor Suspension}
\label{sec:anchor}

As shown in Fig.~\ref{fig:pipeline}, after the 1-pass prediction, we obtain the 1-pass prediction results, which are then refined through anchor suspension. Specifically, after an additional encoding process, we acquire the accurate VMAF and bitrate for prior CRF (i.e. the anchor features). Taking VMAF as an example, if the predicted VMAF for prior CRF is $\text{VMAF}_\text{pred}$, and the true VMAF is $\text{VMAF}_\text{anchor}$, then we can calculate the $\text{offset} = \text{VMAF}_\text{anchor} - \text{VMAF}_\text{pred}$, and apply the offset to shift the entire predicted curve accordingly. Through anchor suspension, we ensure that the predicted curves are accurate at the prior CRF. Therefore, even if the RD performance of the encoder fluctuates slightly due to version updates, our algorithm remains robust with the anchoring of the anchor features.

\subsection{Training Pipeline}
During the training process, we perform two pass of predictions. The 1-pass prediction model takes codec features, content features, and anchor features as input, and outputs 101-dimensional VMAF and 101-dimensional bitrate. Then, we apply anchor suspension to the 1-pass prediction outputs based on the anchor features. The 2-pass prediction model has a similar network structure to the 1-pass prediction model. The target of the 2-pass prediction model is the residual between the anchor-suspended results and the ground truth. In other words, the purpose of the 2-pass model is to fine-tune the anchor-suspended results. Both the two-pass predictions and the anchor suspension operations are differentiable, allowing us to optimize the entire prediction pipeline in an end-to-end manner. The loss function can be written as:
\begin{equation}
 L = \|C_{\text{VMAF}}-G_{\text{VMAF}}\|_2^2 + \lambda * \|C_{\text{Rate}}-G_{\text{Rate}}\|_2^2
\end{equation}
where, $C_{\text{VMAF}}$ and $C_{\text{Rate}}$ denote the predicted results of the rate-quality model, while $G_{\text{VMAF}}$ and $G_{\text{Rate}}$ represent the ground truth. The $\lambda$ is employed to balance the VMAF and bitrate losses, and is experimentally set to $1e-4$.

\begin{figure*}
    \centering
    \subfigure[CRF-VMAF curve]{%
        \includegraphics[width=0.32\textwidth]{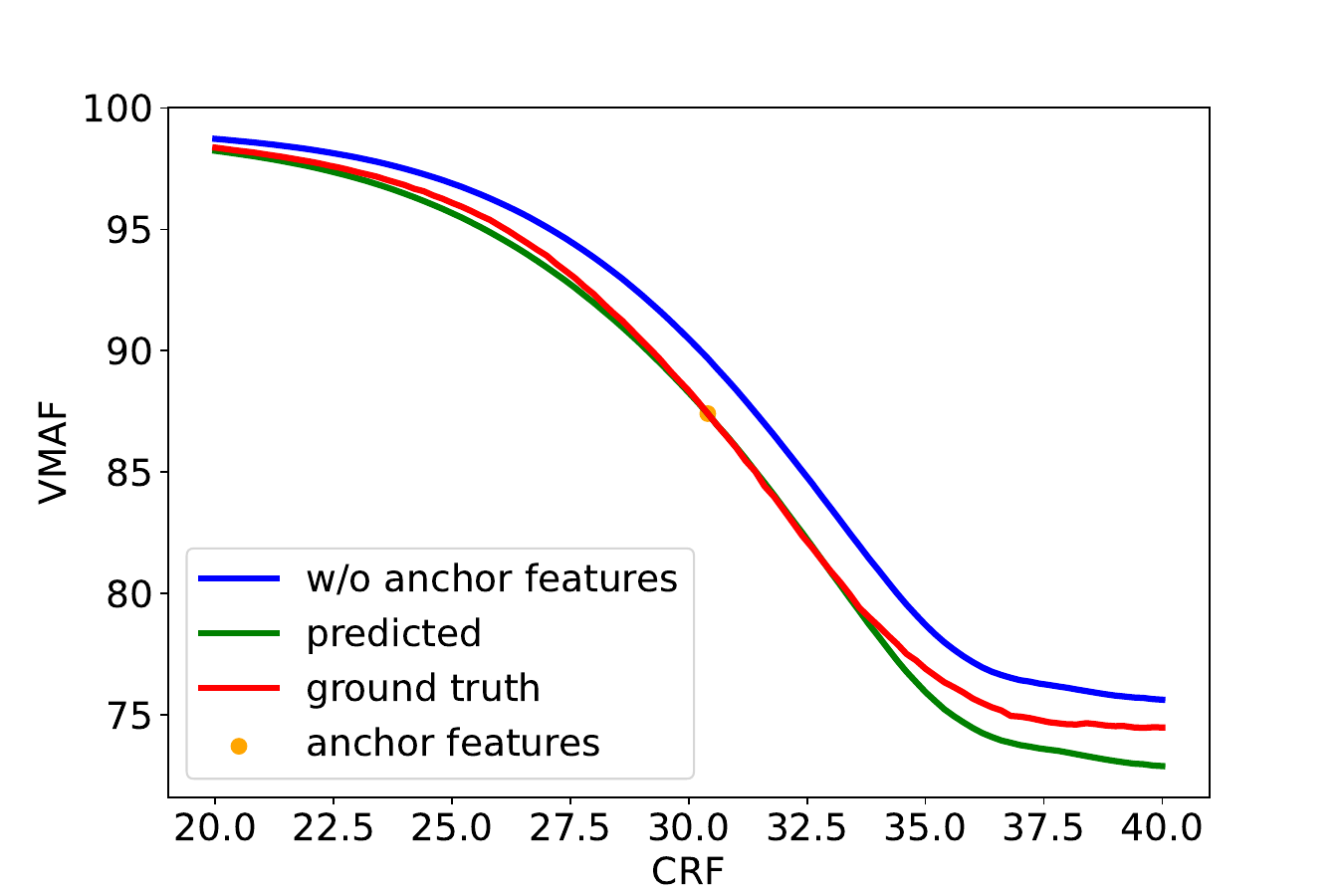}
    }
    \subfigure[CRF-Bitrate curve]{%
        \includegraphics[width=0.32\textwidth]{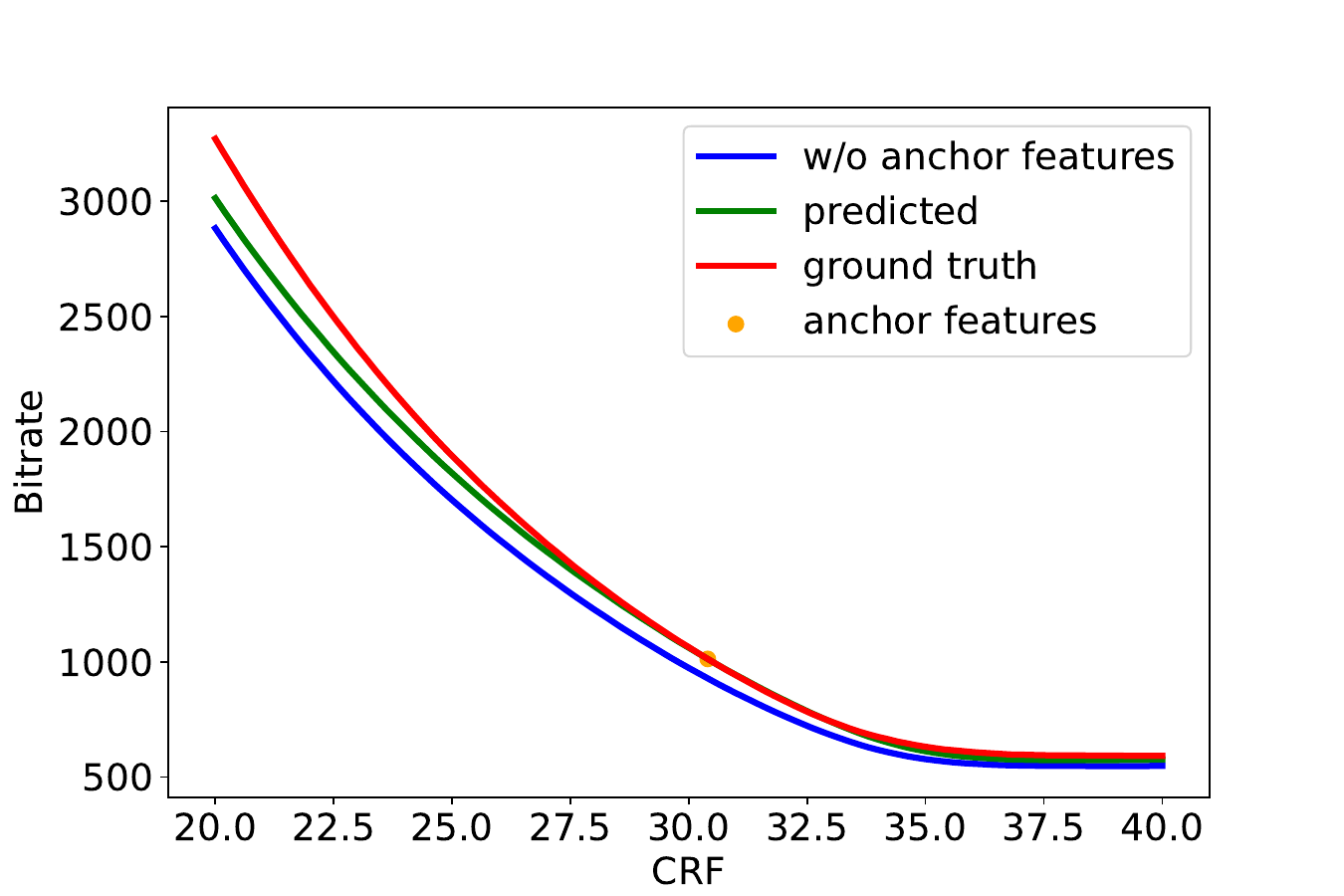}
    }
    \subfigure[Bitrate-VMAF curve]{%
        \includegraphics[width=0.32\textwidth]{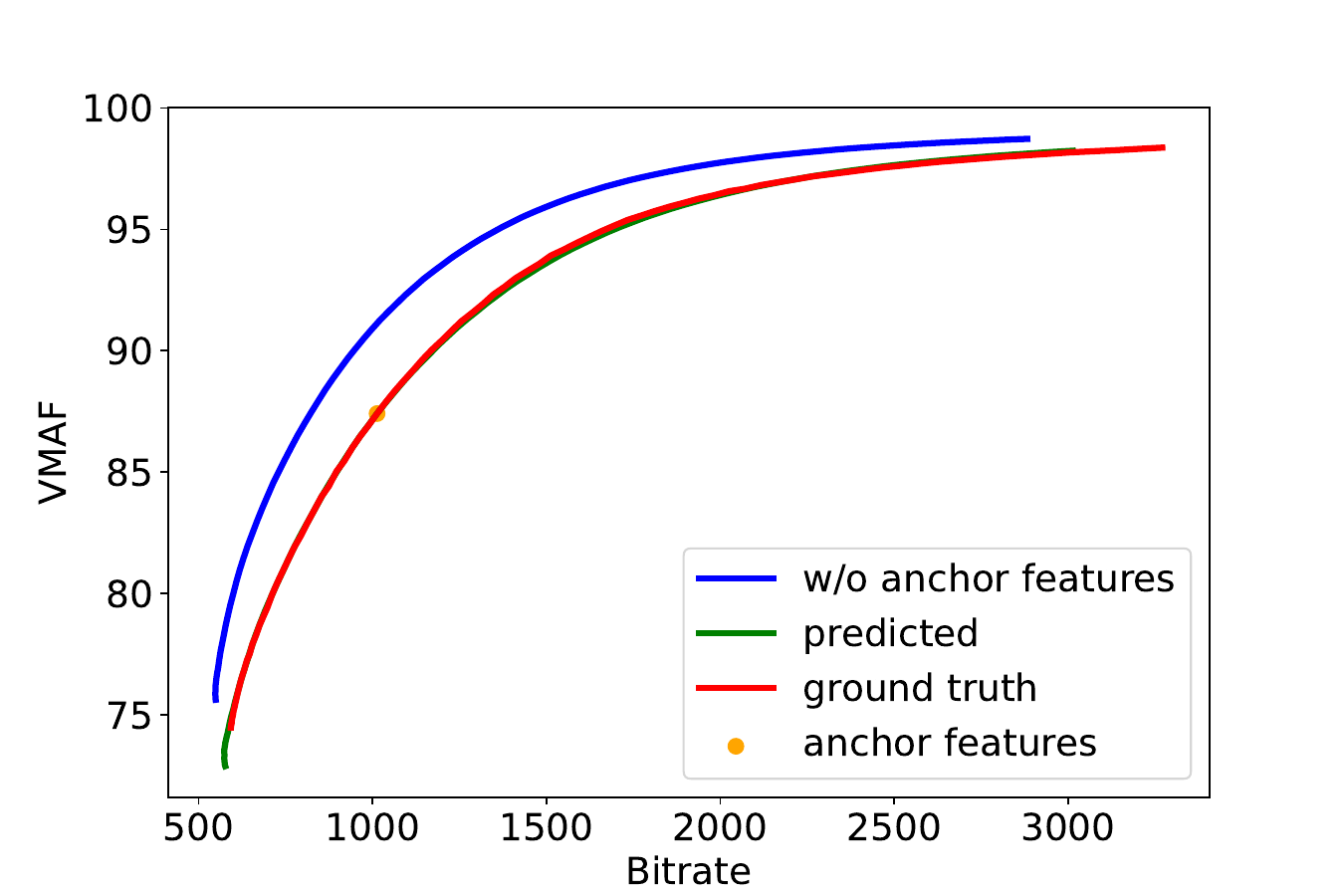}
    }
    \caption{The predicted CRF-VMAF curve, CRF-Bitrate curve, and the corresponding Bitrate-VMAF curve.}
    \label{fig:res}
\end{figure*}

\section{Experiments}
\label{sec:exp}

\subsection{Experimental Setup}

\noindent{\textbf{Data Processing.}} 
During experiments, we conduct the training and validation of the rate-quality model on the red265 encoder. We first collect a dataset of 10,000 online UGC videos, with 8,000 videos as the training data and 2,000 as the test data. To obtain the actual VMAF and bitrate as the training labels, we encode these 10,000 videos using the red265 encoder at 101 CRF points in the range of $[20, 40]$ with a step size of 0.2, and calculate the corresponding VMAF and bitrate. To accelerate the training, we pre-extract the video features (codec features, content features, anchor features) and save them to files. During training, we directly load the features from the files, which improves the training speed, otherwise the feature extraction would become the training bottleneck. Specifically, we use x264 encoder with preset fast to pre-encode the downsampled 360p videos at CRF 18 and CRF 33 to extract the codec features. Additionally, we use red265 encoder to encode the source videos at CRF 30.4 and calculate the VMAF and bitrate as the anchor features.

\vspace{2mm}

\noindent{\textbf{Evaluation Metrics.}}
To evaluate the accuracy of the predicted curves, we calculate the Mean Absolute Error (MAE) of the predicted VMAF curve and bitrate curve as evaluation metrics. Additionally, similar to ~\cite{cai2022quality}, we can predict the CRF corresponding to a target VMAF using the predicted CRF-VMAF curve. Therefore, we also use a metric called VACC, which represents the proportion of videos where the absolute difference between the actual VMAF achieved using the predicted CRF and the target VMAF is less than 1. Mathematically, VACC can be expressed as:
$ \text{VACC} = \frac{1}{N} \sum_{i=1}^{N} \mathbf{1}(d_i < 1)$, where $ d_i = |\text{VMAF}_i - \text{VMAF}_{target}|$. $N$ denotes the total number of videos, $\text{VMAF}_i$ denotes the actual VMAF of the i-th video, and $\mathbf{1}(\cdot)$ is the indicator function, which equals 1 if the condition is true and 0 otherwise. To make a fair comparison with ~\cite{cai2022quality}, we set the target VMAF to 91.

\begin{table}[htb]
\caption{The results of the model prediction accuracy and ablation studies.}
\centering
\vspace{1mm}
\begin{tabular}{c|cc|c}
\toprule[1.2pt]
                     & \multicolumn{2}{c|}{VMAF} & Bitrate(kbps) \\ \cline{2-4} 
                     & MAE        & VACC       & MAE     \\ \hline
w/o anchor features  & 0.716          &    75.33\%          &    87.18     \\
w/o anchor suspension          & 0.300           &  98.34\%            &    51.84     \\
w/o end2end learning & 0.250           & 98.69\%             &    33.31     \\ 
full method & \textbf{0.230}           & \textbf{99.14\%}             &    \textbf{32.25}     \\ 
\bottomrule[1.2pt]
\end{tabular}
\label{tab:ablation}
\end{table}

\begin{table*}[htb]
\caption{Online A/B testing results of a low-quality protect strategy. This result indicates the relative improvement with our rate-quality model over the uniform rate factor method. The square brackets represent the 95\% confidence intervals for online metrics.}
\centering
\begin{tabular}{|c|c|c|c|}
\hline
Video Time                                                                  & Video Views                                                                 & Video Completions                                                           & App Duration                                                                \\ \hline
\begin{tabular}[c]{@{}c@{}}+0.077\%\\ {[}-0.084\%, +0.084\%{]}\end{tabular} & \begin{tabular}[c]{@{}c@{}}+0.107\%\\ {[}-0.087\%, +0.087\%{]}\end{tabular} & \begin{tabular}[c]{@{}c@{}}+0.107\%\\ {[}-0.094\%, +0.094\%{]}\end{tabular} & \begin{tabular}[c]{@{}c@{}}+0.064\%\\ {[}-0.061\%, +0.061\%{]}\end{tabular} \\ \hline
\end{tabular}
\label{ab}
\end{table*}

\subsection{Prediction Accuracy}
As shown in Tab.~\ref{tab:ablation}, the MAE of the VMAF curve predicted by our model is as low as 0.230, and VACC can reach 99.14\%. The MAE of the predicted bitrate curve can be down to 32.25. Fig.~\ref{fig:res} illustrates the predicted CRF-VMAF curve, CRF-bitrate curve, and the corresponding bitrate-VMAF curve for a test video. The results demonstrate that our model's predictions closely align with the ground truth, showcasing the effectiveness of our approach.

The VACC result in ~\cite{cai2022quality} is 98.88\%, while our method can achieve a VACC of \textbf{99.14\%}. Additionally, the training model in ~\cite{cai2022quality} is coupled with the constant-quality encoding strategy, which is beneficial for improving the VACC.
To explore the optimal VACC performance of our method, we try to utilize the training method coupled with the strategy like ~\cite{cai2022quality}. Specifically, for anchor features extraction, we do not use fixed CRF point, but instead found the corresponding CRF for the target VMAF from 1-pass predicted curve, and use this CRF for actual transcoding to obtain the actual bitrate and VMAF. This method resulted in a VACC of \textbf{99.79\%}. However, the model trained in this way is coupled with the strategy and lacks transferability.

\subsection{Ablation Studies}
To validate the effectiveness of each module in the proposed method, we conduct three ablation experiments. \textit{\textbf{w/o anchor features}} denotes that the neural network's input features do not include anchor features, and anchor suspension is also not used. \textit{\textbf{w/o anchor suspension}} indicates that the neural network's input features include anchor features, but anchor suspension is not employed in the subsequent prediction. \textit{\textbf{w/o end-to-end learning}} means that the 1-pass prediction network and 2-pass prediction network are trained separately, without joint end-to-end optimization. Tab.~\ref{tab:ablation} presents the results of the ablation studies, from which we observe that the removal of any module leads to a decline in the prediction accuracy, thereby substantiating the effectiveness of the module we have proposed. Moreover, from Tab.~\ref{tab:ablation}, we can see that the VACC for \textit{\textbf{w/o anchor features}} is 75.33\%, whereas the result in ~\cite{cai2022quality} is only 45.38\%. 
This indicates that even though our task is more challenging in predicting the bitrate-VMAF curve, our superior feature engineering and network architecture design enable us to attain higher accuracy.

\subsection{Online A/B Tests}
Media processing system’s mission is to enhance the video experience for all users. To validate our model, we conducted online experiments to assess whether our rate-quality model enhances user video experience. Over the course of a month, we performed online A/B testing on the media processing system, evaluating performance based on four key metrics: video time, video views, video completions, and app duration. By applying our precise model to improve low-quality videos, the online A/B results, shown in Tab.~\ref{ab}, demonstrated significant improvements. Our strategy achieved increases of +0.107\% in video views, +0.107\% in video completions, and +0.064\% in app duration, all with statistical significance, with a corresponding 2.110\% increase in bitrate cost. These results indicate our model’s effectiveness in enhancing the overall video experience for users.

\section{Disscussion on Encoding Strategy}
Previous CAE method couple the prediction model with the encoding strategy, as changing the target VMAF would require retraining the model. In contrast, our proposed method, which involves predicting the CRF-VMAF curve and CRF-bitrate curve, is decoupled from the encoding strategy. After obtaining the predicted bitrate-VMAF curve, we can freely choose different encoding strategies to maximize the benefit. When constant-quality encoding is required, we can select the corresponding CRF from the CRF-VMAF curve according to target VMAF. 
When maximizing bitrate saving is the priority, we can encode at the CRF with a consistent bitrate-VMAF curve slope, thereby saving bitrate while preserving the visual quality.
Depending on the specific requirements, we can make adjustments without the need to retrain the model.

\section{Conclusion}
In this paper, we propose a rate-quality curve prediction model that is decoupled from the encoding strategy. To reduce the feature extraction cost, we designed a set of video features comprising codec features, content features, and anchor features, which are used to predict the rate-quality curve. Furthermore, we introduced an anchor feature-based anchor suspension method to further enhance the prediction model's performance. Experiments demonstrate that our prediction accuracy is at the current SOTA level in the industry, achieving a VACC of 99.14\%. In addition, we conducted online A/B testing , obtaining both +0.107\% improvements in video views and video completions, and +0.064\% app duration time. 
Based on this prediction model, we can select different encoding strategies to achieve content-adaptive encoding tailored to various requirements, enabling us to optimize for diverse goals such as quality enhancement, bitrate reduction, or a balanced approach.

\section*{Acknowledgment}

This work was supported by the Fundamental Research Funds for the Central Universities, National Natural
Science Foundation of China (62102024, 62331014, 62431015) and Shanghai Key Laboratory of Digital Media Processing and Transmissions, China.

\bibliographystyle{IEEEtran}
\bibliography{refs}

\end{document}